\documentclass[]{report}
\usepackage[utf8]{inputenc}
\usepackage[T1]{fontenc}
\usepackage[french]{babel}
\usepackage{url, fullpage, comment}

\begin{document}

\title{A debate game about societal impacts of AI \\ 
Un jeu-débat sensibilisant aux impacts sociétaux de l'IA}

\author{Carole Adam - MCF, LIG, Univ. Grenoble-Alpes\\ \url{carole.adam@imag.fr} \and Cédric Lauradoux - Univ. Grenoble-Alpes, INRIA Privatics\\ \url{cedric.lauradoux@inria.fr}}

\date{Version auteur - Chapitre à paraître dans l'ouvrage \\"Jouer un rôle pour apprendre, pédagogies en Pratique" \\édité par Hervé Barras et Eric Uyttebrouck \\pour la collection Epistémé chez Eyrolles}

\maketitle
\enlargethispage{35pt}

\small
\section*{Abstract (FR)}
L'Intelligence artificielle (IA) est désormais omniprésente dans nos vies et nous subissons régulièrement ses décisions. Pourtant, la population générale a très peu de connaissances sur son fonctionnement, son usage des données, son manque d'objectivité et sa faillibilité. En accord avec les recommandations de l'Unesco, nous pensons qu'une compréhension basique des algorithmes des modèles d’IA est essentielle pour permettre à tous de choisir en connaissance de cause de les utiliser et de leur confier des données personnelles. Nous proposons pour ce faire un jeu-débat simulant la tenue d’un conseil municipal convoqué pour choisir entre trois solutions d'IA proposées par une entreprise pour faire face à un problème sociétal. Ce dispositif ludique est disponible en téléchargement libre sur Internet. Les résultats des premières sessions montrent son intérêt, non seulement pour la sensibilisation à l'IA, mais aussi pour le développement de compétences d'argumentation et d'écoute.

\section*{Abstract (EN)}
Artificial intelligence (AI) is now ubiquitous in our lives, and we regularly experience its decisions. Yet, the general public has very little knowledge about how it works, its use of data, its lack of objectivity, and its fallibility. In line with UNESCO recommendations, we believe that a basic understanding of AI model algorithms is essential to enable everyone to make informed choices about using them and entrusting them with personal data. To this end, we offer a debate game simulating a municipal council meeting convened to choose between three AI solutions proposed by a company to address a societal problem. This interactive tool is available for free download online. The results of the initial sessions demonstrate its value, not only for raising awareness about AI but also for developing argumentation and listening skills.

\normalsize

\newpage

\chapter{Version française}

\section*{\Large Un jeu-débat sensibilisant aux impacts sociétaux de l'IA \\\strut \\
\large Carole ADAM et Cédric LAURADOUX\\\strut\\
\normalsize Résumé}

L'Intelligence artificielle (IA) est désormais omniprésente dans nos vies et nous subissons régulièrement ses décisions. Pourtant, la population générale a très peu de connaissances sur son fonctionnement, son usage des données, son manque d'objectivité et sa faillibilité. En accord avec les recommandations de l'Unesco, nous pensons qu'une compréhension basique des algorithmes des modèles d’IA est essentielle pour permettre à tous de choisir en connaissance de cause de les utiliser et de leur confier des données personnelles. Nous proposons pour ce faire un jeu-débat simulant la tenue d’un conseil municipal convoqué pour choisir entre trois solutions d'IA proposées par une entreprise pour faire face à un problème sociétal. Ce dispositif ludique est disponible en téléchargement libre sur Internet. Les résultats des premières sessions montrent son intérêt, non seulement pour la sensibilisation à l'IA, mais aussi pour le développement de compétences d'argumentation et d'écoute.

\section{Introduction}
L’intelligence artificielle est de plus en plus présente dans nos vies. Nous lui déléguons nombre de décisions importantes : accès aux formations supérieures, sélection de CV, conduite autonome de véhicules, sécurisation de lieux publics, etc. Bien que soumise aux décisions prises par l’IA, la population générale a très peu de connaissances sur son fonctionnement, et est donc très sensible aux idées véhiculées par les médias : une IA objective, infaillible, qui sauvera le monde ou au contraire le détruira. Par ailleurs, les algorithmes se nourrissent de données personnelles pouvant être détournées ou utilisées à l’encontre des intérêts des personnes (identification d’orientation sexuelle, d’opinions politiques ou religieuses, détection de fréquentation de cliniques d’avortement…).

Nous pensons donc, comme le recommande l’Unesco \cite{unesco}, qu’il est essentiel que la population ait une compréhension basique du fonctionnement des algorithmes des modèles d’IA, pour choisir en connaissance de cause de les utiliser et de leur confier ou non des données. Il s’agit en fait d’apprendre à peser les bénéfices et dangers de toute nouvelle technologie pour faire preuve de discernement face aux promesses des entreprises qui les vendent ou qui utilisent les données produites. Nous proposons un jeu-débat permettant d’évaluer trois solutions d’IA proposées par une entreprise dans le contexte d’une épidémie.

\section{Contexte et problématique}
Dans le contexte de la crise du COVID-19, nous avons souhaité sensibiliser le grand public aux problèmes posés par l’utilisation de l’IA pour la surveillance. Notre but n’est pas d’enseigner les aspects technologiques de l’IA à un public restreint d’informaticiens, mais de faire prendre conscience au plus grand nombre des enjeux sociétaux des nouvelles technologies. En effet, les citoyens sont tous soumis à des décisions prises par des algorithmes d’IA, ce qui pose divers risques (non-respect de la vie privée, décisions biaisées, possibilités de surveillance globale, etc.) dont il faut saisir la portée.

Nous avons travaillé avec l’association de médiation scientifique l’Arbre des Connaissances\footnote{\url{https://arbre-des-connaissances-apsr.org/nos-actions/jouer-a-debattre/}}, qui se donne pour but d’ouvrir la science aux citoyens. L’un de ses outils est une série de jeux-débats sur l’IA et ses applications dans divers domaines, comme les transports ou la santé. L’association apporte des conseils dans l’élaboration du contenu, puis assure l’édition et la diffusion nationale des jeux. Elle propose également des séminaires d’information à destination des enseignants souhaitant les utiliser, et des parcours de formation complets les intégrant.
La série de jeux-débats fixe un cadre global : les participants incarnent des personnages aux rôles prédéfinis qui participent à un conseil municipal, avec un nouvel ordre du jour pour chaque épisode. Dans le cas qui nous intéresse ici, le choix s’est porté sur la surveillance sanitaire en cas d’épidémie. La création d’un nouvel épisode impose certaines contraintes. Tout d’abord, le public cible étant constitué de lycéens, le jeu ne doit nécessiter aucun prérequis : il s’agit donc de bien définir tous les termes spécifiques au thème traité, ici l’IA. La participation est uniquement conditionnée à un niveau de maturité suffisant pour être en mesure de jouer un rôle, prendre du recul, et débattre avec les autres joueurs. Le débat doit pouvoir se jouer sur la durée d’un créneau de cours, soit une heure et demie à deux heures, avec un groupe classe (20 à 30 élèves). Enfin, le jeu doit être facile à animer par les enseignants de lycée, sans leur demander trop de connaissances en IA. Un guide d’animation leur apporte toutes les informations utiles. Le jeu nécessite très peu de matériel et de temps de mise en place, et peut se jouer aussi bien en classe que lors d’événements comme la Fête de la science.

\section{Méthodes}
Les recommandations de l’Unesco \cite{unesco} affirment la nécessité d’éduquer le grand public à l’IA pour lui permettre de prendre des décisions éclairées sur ce qu’il souhaite lui déléguer. Pourtant, seule une petite population de spécialistes est pour l’instant formée et en mesure de comprendre les technologies sous-jacentes. Il existe des MOOCs (Massive Open Online Courses) sur l’Intelligence Artificielle\footnote{MOOC Cnam "IA pour tous" : \url{www.fun-mooc.fr/fr/cours/lintelligence-artificielle-pour-tous/}} ou la protection de la vie privée dans le monde numérique\footnote{MOOC Inria : \url{www.fun-mooc.fr/fr/cours/protection-de-la-vie-privee-dans-le-monde-numerique/}}, mais ce sont des cours plutôt classiques, qui laissent les participants assez passifs. Or, des pédagogies plus actives permettent d’impliquer les apprenants. \cite{simonneaux2001role} montre par exemple l’intérêt d’un jeu-débat pour faire argumenter les participants, en le comparant avec une simple transmission d’informations. \cite{jarvis2002role} utilisent aussi le jeu de rôle comme stratégie d’enseignement, ce qui leur permet de toucher des personnes éloignées de l’informatique.

Plus généralement, \cite{crookall2010serious} définit le jeu sérieux comme un jeu, informatisé ou non, ayant un but autre que ludique, en particulier l’apprentissage. La phase de jeu doit être suivie d’un débriefing permettant de prendre du recul sur ce qui a été appris via le jeu, et de partager les expériences entre joueurs. Des jeux sérieux basés sur la simulation interactive ont par exemple été utilisés pendant la crise sanitaire (\cite{adam2022finding}, \cite{cottineau2020}). De tels outils sont essentiels pour lutter contre la désinformation qui circule, notamment sur les réseaux sociaux, avec des conséquences parfois mortelles en temps d’épidémie, lorsque les fausses nouvelles détournent une partie de la population du vaccin (\cite{lu2022covid}, \cite{nieves2021infodemic}).

Nous souhaitons contribuer à la sensibilisation du plus grand nombre à l’IA, mais aussi à la nécessité de s’informer et de mesurer la balance bénéfice/risque de toute nouvelle technologie. Il a été montré que le jeu peut améliorer l’apprentissage en favorisant l’immersion et l’engagement des joueurs, via un niveau stimulant de difficulté \cite{hamari2016challenging}. Ceci est en accord avec l’approche de la funology \cite{brandtzaeg2018enjoyment}, qui définit l’engagement dans un jeu sérieux selon trois critères issus des travaux de Karasek sur l’engagement au travail~: la demande (le défi proposé au joueur), le contrôle (les moyens mis à sa disposition), et le support (l’interaction avec les pairs). Dans la lignée de ces travaux, nous adoptons l’approche du jeu-débat, qui permet de donner le contrôle au joueur en lui faisant incarner le rôle de conseiller municipal, de lui proposer le défi de concilier des points de vue différents, et de lui offrir un support social, puisque chaque rôle est joué en groupe.

\section{Dispositif réalisé}
\subsection{Phases du jeu}
Nous proposons un jeu de rôle mettant les joueurs en situation de débat municipal \cite{adam2022serious}. Le jeu dure environ une heure et demie, répartie en huit phases (Tableau~\ref{tab1}). La première phase est l’attribution aléatoire des rôles. Elle est suivie de la présentation de l’ordre du jour et des trois solutions proposées. Chaque solution est discutée dans les groupes pour évoquer les avantages et les inconvénients, selon le point de vue de chaque rôle. Les groupes répartissent alors cinq points entre les trois solutions. Les scores, ainsi que les avantages et les inconvénients retenus par les groupes, sont reportés au tableau. L’animateur additionne les points afin de déterminer la solution gagnante. Un débat est ouvert dans l’objectif de modifier les points de vue des groupes. Au terme de ce débat, les groupes peuvent modifier leur répartition de points. Finalement, un débriefing permet de relier cette activité à la vie réelle. Il pointe les technologies déjà utilisées et celles qui pourraient l’être prochainement.

\begin{table}[hbt]
    \centering
    \begin{tabular}{|c|c|}
    \hline
    & \textbf{Phase du jeu} \\
    \hline
    \hline
1 & Attribution aléatoire des rôles aux groupes\\
\hline
2 & Présentation de l’ordre du jour, puis des solutions une par une\\
\hline
3 & Discussion dans les groupes, avantages et inconvénients \\
  & des solutions selon le rôle attribué\\
  \hline
4 & Répartition des points pour chaque solution\\
\hline
5 & Addition des points et report des avantages et inconvénients au tableau\\
\hline
6 & Débat sur les arguments proposés \\
\hline
7 & Modification de la répartition des points\\
\hline
8 & Débriefing de l’activité et institutionnalisation\\
\hline
    \end{tabular}
    \caption{Plan de la séance de jeu-débat}
    \label{tab1}
\end{table}

\subsection{Rôles}
Chaque participant endosse un rôle prédéfini lui permettant de prendre du recul et de changer de perspective. Il y a cinq rôles stéréotypés illustrant une diversité de points de vue. Les Centraux (Central Wafer dans le jeu), ou travailleurs du centre-ville, qui ont des revenus corrects mais peu de temps libre, tiennent à leur liberté, à leur mobilité et celle de leurs clients. Les Alters (Alter Wafer dans le jeu), plutôt jeunes à sensibilité écologiste ou altermondialiste, se déplacent à pied ou à vélo, et demandent des garanties avant d’adopter de nouvelles technologies. Les Seniors sont des personnes âgées, retraitées, vivant en ville ou à la campagne, et focalisées sur le maintien de leur autonomie. Les Futuristes, surtout des jeunes, très connectés et actifs en ligne, adorent l’innovation et les nouvelles technologies, mais ne disposent que de revenus limités. Et enfin les Lointains, un groupe hétérogène, vit en dehors du centre-ville et fait donc de nombreux trajets vers la ville.

Les rôles étant assignés de manière aléatoire, chacun peut être amené à jouer un rôle qui ne correspond pas nécessairement à ses propres opinions. Ceci offre l’avantage d’aider les élèves à prendre de la distance avec l’argumentation et à s’exprimer plus librement. Le débat est dépassionné, car personne n’argumente en son nom propre.

Les participants évaluent d’abord les solutions au sein de leur groupe rôle, constitué de quatre à six élèves selon la taille de la classe. Les arguments sont mis en commun entre tous les groupes dans un deuxième temps.

\subsection{Ordre  du jour et solutions proposées}

Dans le jeu, le conseil municipal est confronté à une problématique, ici une épidémie. Les conseillers doivent y répondre selon les intérêts du rôle qu’ils représentent, tout en assurant le respect de trois critères : préservation de la santé mentale, limitation efficace de la propagation du virus, et équité face aux règles. La société Sowana, spécialisée en IA, propose trois solutions. Eye’Wana consiste en une flotte de caméras de vidéosurveillance et drones, avec reconnaissance faciale, pour détecter et punir automatiquement les violations des règles. Wana’Like est une application de recommandations connectée au profil de réseau social, qui conseille des sorties dans des lieux respectueux du protocole sanitaire et peu fréquentés, en distribuant des coupons de réduction. Wana’Pass propose un passeport sanitaire à points connecté au dossier médical et à des capteurs physiologiques, donnant droit à plus ou moins de sorties selon l’état de santé.

Une seule de ces solutions pourra être adoptée par le conseil municipal qui doit faire un choix. Chaque groupe rôle ayant des intérêts différents, tous vont devoir débattre pour se mettre d’accord.

\subsection{Débats}

Afin de comparer les différentes solutions d’IA, le modérateur guide les participants pour les évaluer à l’aide d’une grille comportant six items (Tableau~\ref{tab2}) inspirés de \cite{Castelluccia2020}. L’efficacité de l’outil est utilisée comme un indicateur de l’équilibre entre les bénéfices et les risques par les agences de protection de la vie privée avant d’autoriser une technologie mobilisant des données personnelles, et par les institutions de santé avant d’autoriser un médicament. L’impact sur les libertés individuelles recherche comment l’outil est contraignant ou restrictif, quelle liberté est sacrifiée, si les bénéfices valent le coût de ce sacrifice, et s’il a des effets secondaires négatifs, par exemple sur la santé mentale. L’impact sur la vie privée définit les modalités de collecte de données, leur degré de sensibilité, la nature des personnes y ayant accès et les conséquences en cas de fuite. Le coût économique et écologique détermine le coût de la technologie, l’identité des financeurs, la pollution générée pour créer, maintenir, utiliser et recycler la solution. L’accessibilité pour tous, et les risques de discrimination vérifient si les différentes solutions assurent une stricte égalité et une équité. Le critère des risques d’erreurs examine les impacts en cas d’erreur, les risques d’échec, de mauvaise utilisation ou de détournement, en pensant à leurs conséquences.

\begin{table}[hbt]
    \centering
    \begin{tabular}{|c|c|}
\hline
    Critère & \textbf{Définition succincte} \\
    \hline
    \hline
Efficacité & Balance bénéfice/risque permettant ou non \\
           & d’autoriser l’utilisation d’une technologie\\
\hline
Impact sur les libertés individuelles & Contraintes, libertés sacrifiées, conséquences \\
& négatives potentielles \\
\hline
Impact sur la vie privée & Collecte des données, accès\\
\hline
Coût économique et écologique & Coût de la technologie et de sa maintenance \\
\hline
Accessibilité, discrimination & Accessible pour tous de manière équitable\\
\hline
Risques d’erreurs & Impact en cas d’erreur, mauvaise utilisation \\
& ou détournement\\
\hline
    \end{tabular}
    \caption{Critères d’évaluation des solutions IA, inspirés de \cite{Castelluccia2020}}
    \label{tab2}
\end{table}

À la fin de la discussion interne, chaque groupe donne les notes qu’il a attribuées à chaque solution. L’animateur reporte au tableau tous les scores ainsi que les principaux arguments. La solution au score total le plus élevé gagne ce premier tour de scrutin. L’animateur déclenche alors un débat entre partisans et adversaires des différentes solutions ; si certains changent d’avis, les scores au tableau peuvent être mis à jour, et la solution choisie peut changer.

\section{Méthodologie}
\subsection{Questionnaire d’évaluation par les joueurs}
Nous avons animé plusieurs ateliers en 2023 pour évaluer le jeu. Une première tentative d'évaluation via un questionnaire en ligne a échoué, car très peu d'élèves ont pris le temps de répondre a posteriori. Nous avons donc ensuite utilisé un questionnaire papier distribué en fin de séance, qui nous a permis de recueillir les retours « à chaud » de tous les participants, soit 81 réponses en cinq séances. Ce questionnaire comporte plusieurs sections : une première concerne le ressenti des joueurs, une deuxième évalue la jouabilité de l’activité, et une dernière concerne son efficacité sur les apprentissages.

Après la collecte des questionnaires sous forme papier, nous avons procédé à leur saisie numérique et au nettoyage des données, afin de permettre leur traitement automatisé, via des scripts Python. Un premier script nous permet de saisir les réponses et de les sauvegarder sous forme structurée. Un deuxième script opère un nettoyage des réponses textuelles pour résoudre les problèmes de casse, d’accentuation, ou d’homogénéisation — les participants ayant pu utiliser des mots différents pour désigner la même solution. Un dernier script permet d'analyser automatiquement ces données structurées (voir résultats en section suivante).

\subsection{Évaluation par les pairs}

Nous avons également présenté ce jeu à des collègues sous diverses formes. Nous avons animé une session test du jeu ; nous l’avons présenté oralement lors d’ateliers pédagogiques universitaires, et lors d’une conférence dédiée aux aspects éthiques de l’IA ; nous l’avons enfin présenté à l’écrit sous forme d’article. Cela nous a permis de recueillir les avis d’une quinzaine d’autres enseignants et/ou chercheurs sur ce jeu et sur la possibilité pour eux de l’utiliser à leur tour.

\section{Résultats, analyse et discussion}

Un premier résultat concernant cette activité est le nombre de téléchargements, soit 231 en juin 2023, un an après la sortie officielle. Malheureusement, nous n’avons pas rédigé de questionnaire à destination des enseignants pour obtenir leur retour sur la facilité d’animer le jeu. Dans ce chapitre, nous analysons deux sources de données : 81 questionnaires d’évaluation collectés auprès des joueurs lors des séances animées par les auteurs d’une part ; des retours qualitatifs de chercheurs et d’enseignants sur les biais de cette activité d’un point de vue pédagogique d’autre part. Les participants n’ont pas forcément tous répondu à toutes les questions du formulaire, aussi précisons-nous pour chaque item de l’analyse le nombre exact de réponses recueillies.

\subsection{Analyse quantitative de la perception du jeu}

Plusieurs questions permettent d’évaluer la perception du jeu. Celui-ci est jugé plutôt intéressant (8/10 de moyenne, sur 58 réponses). Dans les commentaires libres, les lycéens apprécient notamment la possibilité de débattre entre eux. Concernant la durée, 53 la trouvent parfaite, 4 la jugent mal répartie, 5 trouvent le jeu trop court, et 3 le trouvent trop long (65 réponses). Quelques-uns regrettent d’avoir eu trop peu de temps pour débattre à la fin, ou trop pour analyser chaque solution. Cela dépend des préférences individuelles et des dynamiques variables des groupes : un groupe d’opinion homogène convergera rapidement vers une évaluation commune, alors que des divergences internes mèneront à des débats plus longs.

Le thème du jeu-débat concerne la gestion d’une épidémie, en référence à la crise du COVID-19. Parmi les participants (81 réponses), 44 trouvent ce thème pertinent et actuel, 2 le jugent encore trop sensible, tandis que 12 autres le trouvent accessoire. Enfin, 23 répondants ont coché la case « autre ». Parmi eux, 6 seulement ont précisé leur réponse : 4 joueurs disent que le thème, en faisant écho au COVID, est pertinent, tout en notant que cette période de crise leur semble déjà lointaine voire périmée ; 2 trouvent le sujet lassant et expriment un ras-le-bol. Ces réponses sont rassurantes, car peu de joueurs ont été gênés par le thème. Celui-ci n’a pas non plus pénalisé l’engagement dans le jeu : 77 joueurs ont rapporté s’être sentis engagés, contre seulement 2 qui expriment ne pas l’avoir été.

\subsection{Jouabilité et influence des rôles}

Nous avons demandé aux joueurs à quel point ils se reconnaissaient dans le rôle qu’ils avaient dû endosser. La note moyenne d’identification est de 5/10 (81 réponses). Cela montre la diversité des rôles, qui ne correspond pas forcément au profil des joueurs, ainsi que leur caractère stéréotypé, rendant l’identification plus difficile. Cependant, l’objectif des rôles n’est pas que les joueurs s’y identifient, mais qu’ils débattent plus sereinement, tout en comprenant qu’il existe une diversité de points de vue et d’arguments tous différents mais tous raisonnables. Si on s’intéresse à chaque rôle en détail, on constate que l’identification est plus facile avec les Lointains (7/10), probablement car nous avons animé des sessions en zone rurale. Viennent ensuite les Alters (6/10), puis les Centraux (5/10). Étonnamment, les participants se reconnaissent assez peu dans le rôle des Futuristes (5/10). L’identification aux Seniors est enfin la plus faible (4/10).

Nous avons ensuite demandé aux participants s’ils pensaient avoir réussi à jouer leur rôle. La note moyenne obtenue est de 8/10 (sur 81 réponses) : la faible identification à un rôle n’empêche donc pas les joueurs de l’endosser dans le cadre du débat. Les fiches décrivant chaque rôle et le caractère très stéréotypé des différents rôles, ainsi que le recours à des exemples tirés de leur cercle de connaissances, peuvent les y aider. Le rôle n’influe pas significativement sur la capacité ressentie à l’incarner. Paradoxalement, c’est même le rôle des Seniors que les jeunes participants réussissent le mieux à jouer (9/10), probablement car tout le monde a un exemple de grand-parent ou personne âgée dans son entourage. Les autres rôles obtiennent aussi des notes assez hautes : les Lointains, les Centraux et les Alters reçoivent un 8/10, et les Futuristes un 7/10.

Nous avons enfin mesuré l’impact du rôle joué sur l’évaluation du jeu, pour vérifier si les participants bénéficiaient de l’activité quel que soit leur rôle. Les valeurs obtenues dessinent un ordre des rôles les plus bénéfiques : les Futuristes donnent ainsi une moyenne d’engagement de 9/10, d’apprentissage de 8/10, et de réflexion de 8/10, les meilleures notes sur ces trois items. À l’inverse, les Centraux rapportent un intérêt moyen de 7/10, un apprentissage moyen de 6/10, et une réflexion moyenne de 6/10. Entre ces deux extrêmes, tous les groupes notent positivement l’intérêt du jeu   (8/10 pour les Lointains, les Seniors et les Alters) et le niveau de réflexion (8/10 pour les Lointains, 7/10 pour les Alters et les Seniors). Le niveau d’apprentissage est jugé modéré (6/10 pour les trois groupes). L’ordre des rôles est quasiment le même pour chaque item, reflétant une satisfaction inférieure des joueurs ayant incarné les Centraux, les Alters, et dans une moindre mesure les Seniors. Le niveau d’engagement déclaré était le plus faible pour le groupe des Centraux, ce qui peut expliquer que le jeu ait eu moins d’impact sur les joueurs ayant incarné ce rôle, par rapport à des joueurs s’étant sentis plus engagés dans un autre rôle. Cependant, les rôles n’ayant été joués chacun que par une quinzaine de participants, ces résultats sont peu significatifs et peuvent refléter aussi des différences individuelles. Il nous est impossible de le vérifier, notre formulaire ne collectant aucune donnée personnelle sur les joueurs (âge, genre, niveau d’éducation ou connaissances préalables en IA).

\subsection{Accord avec les solutions}

Afin de confirmer cette impression d’avoir bien joué leur rôle, nous avons comparé pour chaque joueur sa solution préférée à titre individuel avec la solution choisie par son groupe. 8 répondants ont indiqué n’avoir aucune solution préférée, les jugeant toutes mauvaises. Parmi les 73 autres réponses, il y en a 49 pour lesquelles le choix du groupe est le même que le choix personnel, contre 24 pour lesquelles il diffère. Certains joueurs ont donc bien voté à l’inverse de leurs préférences personnelles ; pour les autres il est possible que leurs préférences personnelles s’alignent avec celles de leur rôle.

Les participants préfèrent en majorité Wana’Like (37 voix), devant Eye’Wana (18) et Wana’Pass (18). En revanche, à l’issue des séances du jeu, la victoire au conseil municipal revient en premier lieu à Eye’Wana (51 voix, trois séances gagnées) devant Wana’Like (30 voix, deux victoires). Cela confirme que les participants sont capables de défendre une solution au nom du rôle qu’ils jouent, même si elle ne correspond pas à leur opinion personnelle.

\subsection{Analyse qualitative de la perception du jeu}
Nous avons aussi posé des questions plus ouvertes sur le ressenti des joueurs. Les réponses sont en texte libre, et nous y avons analysé les mots les plus fréquents.

Sur le ressenti à l’issue du jeu (57 réponses), le mot qui ressort le plus est « bien » (14 occurrences), suivi de « intéressant » (6) et « réfléchir » (4). Concernant ce que les joueurs ont le plus aimé lors de cette activité (72 réponses), c’est le débat qui ressort – 24 occurrences, dont « débattre » (6), « échanges » (4) et « arguments » (4) –, ainsi que le fait de jouer un rôle au sein d’un groupe – avec « groupe »   (12) et « rôle » (9). Les lycéens semblent beaucoup apprécier de pouvoir discuter entre pairs de la situation sanitaire, et voudraient débattre plus souvent. Les rôles permettent aussi à chacun de prendre du recul et de s’exprimer en toute liberté sans craindre le jugement sur ses opinions personnelles.

Du côté des aspects négatifs, il y a 56 réponses. Différents points ressortent, même si le mot « rien » (7 occurrences) est le plus cité (rien de négatif à signaler). Plusieurs expriment un manque de temps pour les débats. Certains regrettent le peu de connaissances précises obtenues sur l’IA : il faudrait clarifier en introduction que ce n’est pas le but de ce jeu (le but étant de donner des connaissances basiques et un esprit critique sur l’adoption de technologies basées sur l’IA, accessibles au grand public, mais pas d’enseigner les détails techniques du fonctionnement de ces technologies, qui ne seraient alors accessibles qu’avec un certain niveau d’éducation en informatique). D’autres ont mal compris la question et indiquent laquelle des trois solutions ils ont le moins appréciée. Certains joueurs se sentent frustrés que leur solution préférée n’ait pas gagné. Enfin, plusieurs expriment leur frustration car les solutions sont jugées toutes similaires et toutes mauvaises, ou parce qu’il n’y a « pas de bonne réponse~».

Ce dernier point nous semble essentiel. Effectivement, il n’y a pas toujours (voire souvent pas) de « bonne réponse~» à apporter aux problèmes sociétaux. Ce qui est important à enseigner aux élèves, c’est de regarder ces problèmes d’un œil critique, pour se forger leur propre opinion et défendre leurs intérêts. Le côté un peu frustrant du débat nous semble normal et intéressant aussi, car même si tous les participants ont des avis différents, la majorité l’emporte et les autres doivent se plier au choix collectif. Le débriefing est ainsi une partie essentielle d’un tel jeu sérieux, et doit permettre de discuter des émotions ressenties pendant le jeu pour prendre du recul.

\subsection{Analyse quantitative de l’impact subjectif du jeu}

Nous avons ensuite voulu évaluer l’impact du jeu sur les participants. Il est difficile de le mesurer objectivement, car le jeu n’apporte pas de connaissances académiques. Le sujet de l’IA n’est pas au programme au lycée, et relève par conséquent davantage de la culture générale. L’objectif du jeu n’est pas tant d’apporter des connaissances techniques que d’éduquer à un certain esprit critique. Nos questions portaient donc plutôt sur le ressenti des joueurs.

À la question de savoir s’ils pensaient avoir appris des choses lors de l’activité, les participants (54 réponses) ont accordé une note moyenne de 6/10. Cette note assez faible montre bien que le jeu n’apporte pas vraiment de savoirs (« hard skills ») sur l’IA, ce qui peut décevoir, mais plutôt des savoir-être ou compétences sociales (« soft skills »), comme la capacité à débattre, à écouter les arguments d’autrui et à exercer son esprit critique. Le développement de ces compétences est probablement plus difficile à juger pour des lycéens que celui de connaissances concrètes.

Nous leur avons aussi demandé de juger comment le contenu du jeu s’articulait avec leurs études. 29 d’entre eux ont jugé le jeu complémentaire, 9 hors-sujet, 1 redondant, et 42 ont coché « autre », parmi lesquels 7 ont précisé que le sujet était intéressant mais ne faisait pas partie de leur programme. Cela correspond à l’objectif du jeu de donner une culture générale de l’IA à des non-spécialistes.

Une autre question demandait aux joueurs si le jeu les avait fait réfléchir. La note est ici meilleure (7/10 en moyenne, sur 53 réponses), confirmant que l’impact du jeu se situe plus au niveau des réflexions déclenchées que d’un apprentissage scolaire. Pousser les élèves à réfléchir nous semble être essentiel pour l’éducation des citoyens de demain.

Enfin, nous avons interrogé les participants sur ce que le jeu leur avait donné envie de faire par la suite (40 réponses obtenues). Parmi les affirmations proposées (QCM), 18 ont coché « se renseigner sur l’IA », et 20 « poursuivre le débat~», qui étaient nos objectifs principaux. Aucun n’a coché « orienter mes études vers l’IA », ce qui pourrait découler de l’accent mis sur les biais et dangers de l’IA plutôt que sur son fonctionnement technique ou sur ses avantages. Le but n’étant pas de détourner les étudiants de l’IA, il faudrait faire attention à mieux équilibrer le discours lors du jeu.

Le jeu semble donc avoir atteint nos objectifs, même s’il ne s’agit que de déclarations. Les questionnaires étant anonymes et les sessions ayant été organisées dans divers établissements de la région, nous ne sommes pas en mesure d’assurer un suivi à plus long terme pour vérifier si les élèves ont effectivement poursuivi le débat ou la recherche d’informations.

\subsection{Analyse de l’impact objectif du jeu}
Afin d’évaluer plus objectivement l’impact du jeu, nous avons aussi demandé aux participants ce qu’ils en avaient retenu. La réponse est en texte libre, 61 joueurs ont répondu. Plusieurs types de messages émergent~: la nécessaire protection de la vie privée et des données personnelles ; les nombreux biais, dangers, dérives, inconvénients de l’IA, avec des mots comme « dangereuse », « intrusive », « surveillance » ; un sentiment de méfiance, voire de peur (« flippante ») ; et la nécessité de réfléchir, se questionner et se renseigner. Ces messages clés correspondent parfaitement aux centres d’intérêt des deux auteurs de ce chapitre et animateurs des sessions concernées. Il est donc important que les animateurs de ce jeu soient conscients des messages qu’ils veulent faire passer aux participants, et de leurs propres biais éventuels.

\subsection{Analyse des biais du jeu}

Notre objectif à travers cette activité est de sensibiliser les citoyens et d’ouvrir un débat sur les enjeux sociétaux des technologies d’IA utilisées dans notre vie de tous les jours. L’activité ne demande aucune connaissance préalable sur les technologies d’IA, et n’a pas non plus pour but d’en apporter, même si les aspects techniques et éthiques sont intriqués. Certains biais peuvent découler de notre conception de l’activité. Afin de les clarifier, nous avons collecté les retours de chercheurs, enseignants ou ingénieurs pédagogiques lors d’ateliers et conférences.

Le premier point qui ressort est celui des rôles, qui paraissent trop stéréotypés ou peu représentatifs de catégories de population réelles : opinions politiques, catégories socioprofessionnelles, etc. En effet, les rôles sont basés plutôt sur l’âge ou la situation géographique. Cependant, nous tenons à garder des profils non politisés afin de refroidir le débat au lieu de l’échauffer. Par ailleurs, les rôles sont uniquement destinés à montrer une variété de points de vue, sans souci d’exhaustivité. Enfin, notre évaluation montre que les joueurs ont en majorité apprécié de jouer des rôles, qu’ils ont réussi à les incarner même s’ils ne s’y reconnaissaient pas, et qu’ils ont bénéficié du jeu quel que soit le rôle joué.

On nous a aussi reproché la mise en concurrence des rôles lors du débat municipal pour imposer leur solution préférée, au lieu d’une coopération entre rôles pour coconstruire une solution commune. Effectivement, l’activité est centrée sur ce débat d’idées et force chacun à argumenter ses idées face à des points de vue différents. Mais on pourrait prolonger le débat contradictoire par une activité complémentaire d’élaboration conjointe du cahier des charges législatif encadrant l’utilisation de la solution choisie afin de satisfaire tous les citoyens.

Certains regrettent aussi le technosolutionnisme sous-jacent à l’activité, qui force à choisir une des trois solutions proposées. Malheureusement, ce scénario est assez réaliste : la plupart des élèves devront vivre avec de nombreuses technologies d’IA, et doivent donc apprendre à peser leurs bénéfices et risques. Le jeu exige ainsi des joueurs de réfléchir aux points positifs et négatifs de chaque solution avant de prendre une décision, sans jamais les accepter a priori, et d’argumenter tous leurs choix. Ces compétences leur seront très utiles dans leur vie de citoyens.

Enfin, on peut noter que les technologies proposées n’ont de sens que dans un certain cadre législatif, leurs conséquences sur les libertés individuelles dépendant de leur utilisation. Cependant, dans le cadre du jeu, ce contexte légal est volontairement assez flou, afin de conserver une certaine généricité. En effet, les mêmes questions doivent selon nous se poser quel que soit le contexte, car une technologie mise en place pour un objectif particulier et dans un cadre légal donné peut ensuite plus facilement être pérennisée et étendue, et le cadre législatif peut lui-même évoluer. Il est donc important pour les citoyens de se poser des questions, d’avoir conscience des possibles détournements d’une technologie et de fixer des garde-fous, comme une durée maximale d’utilisation. Ainsi, lors du débriefing, les animateurs fournissent des informations sur les protections légales pour les utilisateurs de technologies d’IA, en particulier le droit, garanti par le Règlement général sur la protection des données mais souvent ignoré, de refuser qu’une solution d’IA prenne automatiquement des décisions les concernant.

\section{Conseils de mise en œuvre pratique}

Ce jeu-débat est publié sur le site de l’Arbre des Connaissances depuis avril 2022, dans le cadre de leur série de jeux-débats consacrés à l’IA. Les différents épisodes sont téléchargés et animés régulièrement en classe par des enseignants. La mise en œuvre est facile grâce aux éléments fournis et aux quelques conseils ci-dessous.

\subsection{Public cible}
L’Arbre des Connaissances développe ses activités spécifiquement pour les lycées. Cependant, nous avons testé notre jeu-débat également avec des étudiants. Il a semblé peu adapté à des élèves de niveau master ayant déjà des connaissances précises sur l’IA, et attendant du contenu plus technique. En revanche, il fonctionne parfaitement en licence, où au-delà du thème de l’IA, il permet surtout d’apporter des compétences de débat, d’argumentation, d’esprit critique, qui seront très utiles pour les études, quel que soit leur parcours. Il peut aussi servir à former des enseignants non informaticiens  . Enfin, le thème peut être adapté pour porter une réflexion et un débat entre enseignants sur la place croissante de l’IA dans l’éducation.

\subsection{Mise en place de l’activité}
Pour mettre en œuvre ce jeu en classe, il faut disposer d’un créneau d’environ une heure et demie, et d’une classe constituée d’entre 15 et 30 élèves. Afin de faciliter les discussions à l’intérieur des groupes, on pourra organiser les tables en cinq îlots séparés. Un grand tableau permettra de reporter les évaluations avant le débat. Des feutres ou craies de plusieurs couleurs seront utiles si l’on veut procéder à plusieurs phases de débat, afin de visualiser l’évolution des scores.

Toutes les pièces nécessaires sont disponibles librement sur inscription sur le site de l’Arbre des Connaissances. Il suffit de les télécharger et de les imprimer. Le fichier contient les cartes décrivant les cinq rôles, identiques dans tous les épisodes de la série. Il faut prévoir une carte par rôle, à poser à l’envers sur la table du groupe concerné. La répartition dans les groupes se fait au hasard afin d’éviter que chacun choisisse un groupe dans lequel il se reconnaît mieux.

Le fichier contient ensuite une fiche expliquant l’ordre du jour du conseil municipal, spécifique à l’épisode (ici, faire face à une épidémie). Cette fiche permet de mettre l’accent sur les points importants à prendre en compte lors de l’évaluation des solutions. Une seule fiche est nécessaire, l’animateur se chargeant de la lire et de l’expliquer en début de séance.
Enfin, le fichier contient les fiches solutions, elles aussi spécifiques à chaque épisode. Il y a une fiche par solution, qui fournit à la fois l’argumentaire marketing de Sowana, quelques détails sur les techniques d’IA utilisées dans cette solution, et le témoignage d’un citoyen destiné à soulever quelques premières interrogations. Ces fiches informent les joueurs sur chaque solution sans être trop techniques, et en suggèrent quelques limites à travers un témoignage d’utilisation, afin d’amorcer le débat interne au groupe. Il faut imprimer chaque fiche solution en cinq exemplaires (un par groupe), et les distribuer l’une après l’autre, en laissant le temps du débat interne sur chaque solution avant de distribuer la suivante.

\subsection{Guide de l’animation}

Le site de l’Arbre des Connaissances fournit des liens vers des ressources supplémentaires ainsi qu’un livret guide de l’animateur, destiné aux professeurs qui superviseront le jeu. Ce guide contient des définitions utiles et des informations plus techniques pour approfondir les notions en cours ; des questions et échelles d’évaluation pour alimenter le débat ; des informations sur des solutions d’IA similaires déjà réelles à présenter lors du débriefing ; ainsi que des liens vers de la documentation supplémentaire pour aller plus loin sur le sujet.

Le rôle de l’animateur est essentiel au bon déroulement de l’activité. C’est lui qui modère le débat, pour assurer comme recommandé par \cite{dieleman2006games} un déroulement juste, serein et équilibré. Il peut aussi être amené à rappeler aux participants de bien parler au nom de leur rôle et pas en leur opinion propre. L’animateur lui-même doit lui aussi veiller à rester neutre, et guider le débat sans l’orienter ; il doit être conscient de ses propres biais et du risque de les communiquer aux joueurs.

L’objectif de ce jeu est d’éduquer au sujet de l’IA, d’un point de vue technique (fonctionnement général des algorithmes sous-jacents), mais aussi éthique (questions soulevées par son utilisation). Le simple fait de jouer et de débattre n’est cependant pas suffisant pour atteindre cet objectif. À la fin de l’activité, l’animateur a donc la charge de débriefer le jeu pour permettre l’assimilation des messages. Un bon débriefing est indispensable à l’apprentissage via un jeu sérieux (\cite{crookall2010serious} ; \cite{dieleman2006games} ; \cite{whalen2018all}) et poursuit plusieurs buts : clarifier les concepts manipulés pendant la session de jeu ; relier les concepts simplifiés pendant le jeu avec la complexité du monde réel, pour assurer le transfert des apprentissages en situation réelle ; et partager les expériences, les émotions et la réflexion entre les participants. Le débriefing doit donc être conduit par un modérateur informé, avec à l’esprit un objectif pédagogique clairement défini.

Dans le cadre de ce jeu-débat sur l’utilisation de l’IA pour la surveillance sanitaire, l’animateur pourra montrer comment les différentes solutions évoquées pendant le jeu renvoient en fait à des solutions bien réelles déjà mises en place dans certains pays. Ces exemples illustrent très concrètement les risques d’une dérive de l’usage utile à l’usage forcé de l’IA, et aident les participants à comprendre notre message clé : ils sont tous concernés par l’IA et son utilisation.

\section{Conclusion} \enlargethispage{15pt}
Dans ce chapitre, nous avons présenté un jeu-débat permettant de discuter avec des élèves des bénéfices et risques des technologies d’IA appliquées à un problème sociétal, ici la gestion d’une épidémie. Le thème spécifique est accessoire, les questions soulevées étant bien plus générales. L’IA est en effet de plus en plus prégnante et tous les citoyens doivent être capables de prendre des décisions éclairées à son sujet. Par ailleurs, et au-delà même de l’IA, les participants ont surtout apprécié de pouvoir débattre entre pairs. Les compétences apportées par ce jeu concernent l’argumentation, l’esprit critique, et l’écoute de points de vue variés, des savoir-être essentiels pour les futurs citoyens. À ce titre, le principe de ce jeu-débat pourrait être étendu à d’autres thèmes et disciplines.

\small \enlargethispage{35pt}
\bibliographystyle{abbrv}

\newpage
\subsection*{Annexe : questionnaire détaillé}
Voici la liste complète des questions posées : 
\small{\texttt{
\begin{itemize}
    \item Quel groupe représentiez-vous ? (case à cocher)
    \item Vous reconnaissez-vous dans ce rôle ? De O (pas du tout, j'ai dû faire un effort pour me mettre dans la peau de mon personnage) à 10 (complètement, les arguments que j'ai utilisés correspondent complètement à mon avis personnel)
    \item Avez-vous réussi à jouer ce rôle ? De 0 (non, j'ai plutôt donné mon opinion personnelle) à 10 (oui, j'ai complètement endossé mon personnage)
    \item Combien de points votre groupe a-t-il attribué à chaque solution ?
    \item Quelle solution a été sélectionnée par votre conseil municipal ? (case à cocher)
    \item Quelle solution auriez-vous préféré à titre personnel ? (case à cocher + justification libre)
    \item Avez-vous changé votre opinion personnelle sur les solutions proposées pendant le jeu ? (case à cocher + explication libre)
    \item Comment vous sentez-vous à l'issue du jeu ?
    \item Avez-vous eu le sentiment d'être engagé dans le jeu ? (oui/non ou valeur de 0 à 10 selon les versions du questionnaire + explication libre)
    \item Qu'avez-vous le plus aimé ?
    \item Qu'avez-vous le moins aimé ?
    \item Quels messages clés retenez-vous du jeu ?
    \item Que pensez-vous de la durée du jeu ? Case à cocher parmi : trop court, parfait, trop long, mal réparti
    \item Pensez-vous que d'autres rôles devraient être représentés dans le jeu ? Si oui lesquels ?
    \item Pensez-vous à d'autres solutions possibles mettant à profit l'IA pour lutter contre la crise sanitaire? Lesquelles ?
    \item Que pensez-vous du thème du jeu (crise sanitaire) ? Case à cocher parmi : pertinent et actuel, encore trop sensible, peu importe le thème, ou autre (précisez).
    \item Ce jeu est-il intéressant ? de 0 (pas du tout) à 10 (passionnant)
    \item Complémentarité avec les autres cours de la formation ? Case à cocher parmi : redondant, complémentaire, hors-sujet, autre (précisez). 
    \item Avez-vous appris des choses ? de 0 (rien du tout) à 10 (oui beaucoup de choses) + précisions libres
    \item Est-ce que ce jeu vous a fait réfléchir sur l'IA ? de 0 (pas du tout) à 10 (oui beaucoup) + précisions libres
    \item Est-ce que ce jeu a déclenché des débats entre collègues après la séance ? Oui / Non + précisions libres
    \item Le jeu vous a donné envie de : poursuivre le débat ensuite, vous renseigner sur l'IA, orienter vos études vers l'IA, autre (précisez). (question ajoutée après la première séance de 15 participants)
\end{itemize}}

\normalsize

\newpage

\chapter{English version}
\section*{\Large A debate game to raise awareness about societal impacts of AI\\\strut \\
\large Carole ADAM and Cédric LAURADOUX\\\strut\\
\normalsize Abstract}

Artificial intelligence (AI) is now ubiquitous in our lives, and we regularly experience its decisions. Yet, the general public has very little knowledge about how it works, its use of data, its lack of objectivity, and its fallibility. In line with UNESCO recommendations, we believe that a basic understanding of AI algorithms is essential to enable everyone to make informed choices about using them and entrusting them with personal data. To this end, we offer a debate game simulating a municipal council meeting convened to choose between three AI solutions proposed by a company to address a societal problem. This tool is available for free download online. The results of the initial sessions demonstrate its value, not only for raising awareness about AI but also for developing argumentation and listening skills in students.

\section{Introduction}
Artificial intelligence is increasingly present in our lives. We delegate many important decisions to it: access to higher education, CV screening, autonomous vehicle driving, securing public spaces, etc. Although subject to decisions made by AI, the general population has very little knowledge of how it works and is therefore very sensitive to the ideas conveyed by the media: an AI that is objective, infallible, and will either save the world or, on the contrary, destroy it. Furthermore, algorithms feed on personal data that can be misappropriated or used against individuals' interests (identification of sexual orientation, political or religious opinions, detection of visits to abortion clinics, etc.).

We therefore believe, as recommended by UNESCO \cite{unesco}, that it is essential for the general population to have a basic understanding of how AI algorithms work in order to make informed choices about using them and whether to entrust them with personal data or not. It is actually necessary to learn to weigh the benefits and dangers of any new technology to exercise discernment in the face of promises from companies that sell them or use the data produced. To promote such discernment, we propose a debate game where players have to evaluate three AI solutions offered by a company in the context of an epidemic.

\section{Context and Problem Statement}

In the context of the COVID-19 crisis, we aimed to raise public awareness regarding the issues raised by the use of AI for sanitary surveillance. Our goal is not to teach the technological aspects of AI to a restricted audience of computer scientists, but to make as many people as possible aware of the societal stakes of new technologies. Indeed, all citizens are subject to decisions made by AI algorithms, which poses various risks (lack of privacy, biased decisions, possibilities for global surveillance, etc) whose scope must be understood.

We worked with the French scientific mediation association "L’Arbre des Connaissances"\footnote{\url{https://arbre-des-connaissances-apsr.org/nos-actions/jouer-a-debattre/}}, which aims to open science to citizens. One of its tools is a series of debate games on AI and its applications in various fields, such as transport or health. The association provides advice on content development, then handles the publishing and national distribution of the games. It also offers information seminars for teachers wishing to use them, as well as complete training programs that incorporate them.

The debate-game series establishes a global framework: players embody characters with predefined roles, participating in a city council meeting with a new agenda for each episode. In our episode, the agenda focuses on health surveillance during an epidemic. Creating a new episode imposes certain constraints. First, as the target audience consists of high school students, the game must not impose any prerequisites; it is therefore necessary to clearly define all technical terms related to AI. Participation is only conditioned by a level of maturity sufficient to be able to play a role, step back, and debate with other players. The debate must be playable within the duration of a class period, \emph{i.e.} one and a half to two hours, with a class group (20 to 30 students). Finally, the game must be easy to animate for high school teachers of any discipline, without requiring extensive knowledge of AI on their part either. A facilitation guide provides them with all the useful information. The game requires very little material or setup time and can be played in class as well as during events like the "Fête de la Science".

\section{Methods}
UNESCO's recommendations \cite{unesco} assert the need to educate the general public about AI to enable them to make informed decisions about what they wish to delegate to it. However, only a small population of specialists is currently trained and able to understand the underlying technologies. There exist various MOOCs (Massive Open Online Courses) on Artificial Intelligence\footnote{Cnam MOOC "AI for all": \url{www.fun-mooc.fr/fr/cours/lintelligence-artificielle-pour-tous/}} or privacy protection in the digital world\footnote{Inria MOOC: \url{www.fun-mooc.fr/fr/cours/protection-de-la-vie-privee-dans-le-monde-numerique/}}, but these are rather traditional courses that leave participants quite passive. Yet, active pedagogies allow for greater learner involvement. For example, \cite{simonneaux2001role} shows the value of a debate-game in encouraging participants to argue, comparing it with a simple transmission of information. \cite{jarvis2002role} also use role-playing as a teaching strategy,  which allows them to reach people with no computer science skills.

More generally, \cite{crookall2010serious} defines a serious game as a game, computerized or not, having a purpose other than entertainment, specifically learning. The game phase must be followed by a debriefing to reflect on what was learned through the game and to share experiences between players. Serious games based on interactive simulation were used, for example, during the health crisis (\cite{adam2022finding}, \cite{cottineau2020}). Such tools are essential to counter the circulating misinformation, particularly on social media, which can have fatal consequences during an epidemic when fake news dissuades part of the population from vaccinating (\cite{lu2022covid}, \cite{nieves2021infodemic}).

We wish to contribute to raising awareness among the widest possible audience about AI as well as about the need to stay informed and to weigh the benefit-risk balance of any new technology. It has been shown that gamification can improve learning by promoting immersion and player engagement through a stimulating level of difficulty \cite{hamari2016challenging}. This is in line with the "funology" approach \cite{brandtzaeg2018enjoyment}, which defines engagement in a serious game according to three criteria derived from Karasek's work on job engagement: demand (the challenge offered to the player), control (their resources and autonomy to face demands), and social support (interaction with peers). Following this line of work, we adopt the debate game approach, which allows us: to give \textbf{control} to the player by having them play the role of a city councillor; to offer them the \textbf{challenge} of reconciling different points of view; and to provide them with social \textbf{support} since each role is played in a group.

\section{Designed game}

\enlargethispage{35pt}
\subsection{Game Phases}
We propose a role-playing game that places players in a municipal debate situation \cite{adam2022serious}. The game lasts approximately one and a half hours, divided into eight phases (Table~\ref{tab1}). The first phase is the random assignment of roles. This is followed by the presentation of the agenda and the three proposed solutions. Each solution is discussed within groups to highlight its advantages and drawbacks from the perspective of each role. Groups then distribute five points among the three solutions. The scores, as well as the advantages and drawbacks identified by the groups, are written on the board. The facilitator adds up the points to determine the winning solution. A debate is then opened with the aim of shifting the groups' perspectives. At the end of this debate, groups may modify their points allocation which might change the winning solution. Finally, a debriefing session links this activity to real life, highlighting technologies already in use and those that could be implemented soon.

\begin{table}[hbt]
    \centering
    \begin{tabular}{|c|c|}
    \hline
    & \textbf{Game Phase} \\
    \hline
    \hline
1 & Random assignment of roles to groups\\
\hline
2 & Presentation of the agenda, then the solutions one by one\\
\hline
3 & Group discussion: advantages and disadvantages \\
  & of the solutions according to the assigned role\\
  \hline
4 & Distribution of points for each solution\\
\hline
5 & Totalling of points and recording of advantages and drawbacks on the board\\
\hline
6 & Debate on the proposed arguments \\
\hline
7 & Modification of the point distribution\\
\hline
8 & Activity debriefing and institutionalisation\\
\hline
    \end{tabular}
    \caption{Plan of the debate-game session}
    \label{tab1}
\end{table}

\subsection{Roles}
Each participant takes on a predefined role, allowing them to step back and shift their perspective. There are five stereotyped roles illustrating a diversity of viewpoints. The \textit{Centrals} (Central Wafer in the game), or downtown workers, have decent incomes but little free time; they value their freedom, their mobility, and that of their customers. The \textit{Alters} (Alter Wafer in the game) are mostly young people with ecological or anti-globalization leanings; they travel on foot or by bicycle and demand guarantees before adopting new technologies. The \textit{Seniors} are elderly, retired individuals living in the city or the countryside, focused on maintaining their autonomy. The \textit{Futurists}, mainly young people, are highly connected and active online; they love innovation and new technologies but have limited income. Finally, the Remote (\textit{Lointains}) are a heterogeneous group living outside the city centre and making frequent commutes to the city.

Since roles are assigned randomly, everyone may be required to play a role that does not necessarily match their own opinions. This offers the advantage of helping students distance themselves from the arguments and express themselves more freely. The debate is dispassionate because no one is arguing in their own name.

Participants first evaluate the solutions within their role group, consisting of four to six students depending on the class size. Arguments are then shared among all groups in a second phase.

\subsection{Agenda and Proposed Solutions}

In the game, the city council is faced with a problem, in this case an epidemic. The councillors must respond according to the interests of the role they represent, while ensuring compliance with three criteria: preservation of mental health, effective limitation of the virus spread, and fairness regarding the rules. The company \textit{Sowana}, specialized in AI, proposes three solutions. \textit{Eye’Wana} consists of a fleet of CCTV cameras and drones with facial recognition to automatically detect and punish violation of sanitary rules. \textit{Wana’Like} is a recommendation app connected to social media profiles; it promotes outings to uncrowded places that follow the health protocol, by distributing discount coupons. \textit{Wana’Pass} proposes a points-based health passport connected to medical records and physiological sensors, granting more or fewer outings depending on one's health status.

Only one of these solutions can be adopted by the city council, which must make a choice. Since each role group has different interests, they will have to debate together in order to reach an agreement.

\subsection{Debates}

To compare the different AI solutions, the moderator guides participants in evaluating them along a grid consisting of six items (Table~\ref{tab2}) inspired by \cite{Castelluccia2020}. The effectiveness of the tool is used as an indicator of the balance between benefits and risks by privacy protection agencies before authorizing a technology involving personal data, and by health institutions before authorizing a new drug. The impact on individual liberties examines how the tool is coercive or restrictive, which liberty is sacrificed, whether the benefits are worth the cost of this sacrifice, and whether it has negative side effects, for example on mental health. The impact on privacy defines the methods of data collection, their degree of sensitivity, the nature of the people who can access them, and the consequences in the event of a leak. The economic and ecological cost determines the cost of the technology, the identity of the funders, and the pollution generated to create, maintain, use, and recycle the solution. Accessibility for all and the risks of discrimination check whether the various solutions ensure strict equality and equity. The error risk criterion examines the impacts in case of error, the risks of failure, misuse, or misappropriation, while considering their consequences.

\begin{table}[hbt]
    \centering
    \begin{tabular}{|c|c|}
\hline
    Criterion & \textbf{Brief Definition} \\
    \hline
    \hline
Effectiveness & Benefit/risk balance determining whether or not \\
               & the use of a technology is authorized\\
\hline
Impact on individual liberties & Constraints, sacrificed liberties,\\
& potential negative consequences \\
\hline
Impact on privacy & Data collection, access\\
\hline
Economic and ecological cost & Cost of technology and its maintenance \\
\hline
Accessibility, discrimination & Accessible to all in an equitable manner\\
\hline
Risk of errors & Impact in case of error, misuse, or misappropriation\\
\hline
    \end{tabular}
    \caption{Evaluation criteria for AI solutions, inspired by \cite{Castelluccia2020}}
    \label{tab2}
\end{table}

At the end of the internal discussion, each group provides the scores they assigned to each solution along with some justification for these scores. The facilitator records all scores and the main arguments on the board. The solution with the highest total score wins this first round of voting. The facilitator then initiates a debate between supporters and opponents of the various solutions; if some groups change their minds, the scores on the board can be updated, and the chosen solution may change.

\section{Methodology}

\subsection{Player Evaluation Questionnaire}
We animated several workshops in 2023 to evaluate the game. An initial attempt at evaluation via an online questionnaire failed, as very few students took the time to respond after the session. We therefore subsequently used a paper questionnaire distributed at the end of the session, which allowed us to collect "on-the-spot" feedback from all participants, resulting in 81 responses across five sessions. This questionnaire includes several sections: the first concerns the players' feelings, the second evaluates the playability of the activity, and the last assesses its effectiveness on learning.

After collecting the paper questionnaires, we proceeded with digital entry and data cleaning to allow for automated processing via Python scripts. A first script allowed us to enter the responses and save them in a structured format. A second script performed cleaning of the textual responses to resolve issues with case, accents, or homogenization, since participants may have used different words to refer to the same solution. A final script enabled the automatic analysis of this structured data (see results in the following section).

\subsection{Peer evaluation}
We also presented this game to colleagues in various forms. We facilitated a test session of the game; we presented it orally during university pedagogy workshops and at a conference dedicated to the ethical aspects of AI; finally, we presented it in writing as an article. This allowed us to collect the opinions of about fifteen other teachers and/or researchers regarding this game and the possibility for them to use it in turn.

\section{Results, Analysis, and Discussion}

A primary result regarding this activity is the number of downloads: 231 as of June 2023, one year after the official release. Unfortunately, we did not design a questionnaire for teachers to obtain their feedback on the ease of facilitating the game. In this chapter, we analyse two sources of data: 81 evaluation questionnaires collected from players during sessions facilitated by the authors; and qualitative feedback from researchers and teachers regarding the pedagogical biases of this activity. Not all participants necessarily answered every question on the form, so we specify the exact number of responses collected for each analysis item.

\subsection{Quantitative Analysis of Game Perception}

Several questions allow for the evaluation of the game perception. It is judged as quite interesting (average of 8/10, out of 58 responses). In the open comments, high school students particularly appreciate the opportunity to debate among themselves. Regarding the duration, 53 found it perfect, 4 judged it poorly distributed, 5 found the game too short, and 3 found it too long (65 responses). A few participants regretted having too little time to debate at the end or too much time to analyse each solution. This depends on individual preferences and varying group dynamics: a group with a homogeneous opinion will quickly converge on a common evaluation, whereas internal disagreements will lead to longer debates.

The theme of the debate game concerns the management of an epidemic, referencing the COVID-19 crisis. Among the participants (81 responses), 44 found this theme relevant and current, 2 judged it still too sensitive, while 12 others found it incidental. Finally, 23 respondents checked the "other" box. Among them, only 6 clarified their response: 4 players stated that the theme echoing COVID-19 was relevant, while noting that this crisis period already feels distant or even outdated to them; 2 found the subject tiresome and expressed a sense of fatigue (feeling "fed up"). These responses are reassuring, showing that few players were bothered by the theme. It also did not hinder engagement in the game: 77 players reported feeling engaged, compared to only 2 who expressed that they were not.

\subsection{Playability and Influence of Roles}

We asked players to what extent they recognized themselves in the role they had to assume. The average identification score was 5/10 (81 responses). This reflects the diversity of the roles, which do not necessarily match the players' profiles, as well as their stereotyped nature, making identification more difficult. However, the objective of the roles is not for players to identify with them, but rather to debate more calmly, while understanding that there exist a diversity of viewpoints and arguments, all different yet all reasonable. Looking at each role in detail, identification was easiest with the \textit{Remote} (7/10), likely because we facilitated many sessions in rural areas. This was followed by the \textit{Alters} (6/10), then the \textit{Centrals} (5/10). Surprisingly, participants identified relatively little with the \textit{Futurist} role (5/10). Finally, identification with the \textit{Seniors} was the lowest (4/10).

We then asked participants if they believed they had succeeded in playing their role. The average score obtained was 8/10 (out of 81 responses). Thus, a low identification with a role does not prevent players from embodying it within the context of the debate. The fact sheets describing each role, the highly stereotyped nature of the characters, and the use of examples from their own social circles as role models may have helped them. The role does not significantly influence the perceived ability to personify it. Paradoxically, it was the \textit{Senior} role that young participants managed to play best (9/10), likely because everyone has an example of a grandparent or an elderly person in their family circle. Other roles also received fairly high scores: \textit{Remotes}, \textit{Centrals}, and \textit{Alters} received an 8/10, and \textit{Futurists} a 7/10.

Finally, we measured the impact of the assigned role on the game evaluation to verify whether participants benefited from the activity regardless of their role. The values obtained outline a ranking of the most beneficial roles: the \textit{Futurists} gave an average engagement score of 9/10, a learning score of 8/10, and a reflection score of 8/10, the highest marks across these three items. Conversely, the \textit{Centrals} reported an average interest of 7/10, average learning of 6/10, and average reflection of 6/10. Between these two extremes, all groups rated the game's interest positively (8/10 for the \textit{Remotes}, \textit{Seniors}, and \textit{Alters}) as well as the level of reflection (8/10 for the \textit{Remotes}, 7/10 for the \textit{Alters} and \textit{Seniors}). The level of learning was judged moderate (6/10 for all three groups). The ranking of the roles remains nearly identical for each item, reflecting lower satisfaction among players who portrayed the \textit{Centrals}, the \textit{Alters}, and to a lesser extent the \textit{Seniors}. The self-reported engagement level was lowest for the \textit{Central} group, which may explain why the game had less impact on players portraying this role compared to those who felt more engaged in another. However, as each role was played by only about fifteen participants, these results are not statistically significant and may also reflect individual differences. It is impossible for us to verify this, as our form did not collect any personal data on the players (age, gender, education level, or prior knowledge of AI).

\subsection{Agreement with the Solutions}

To confirm the impression that roles were portrayed effectively, we compared each player's individual favourite solution with the solution chosen by their group. Eight respondents indicated that they had no preferred solution, judging them all bad. Among the other 73 responses, there were 49 for whom the group's choice matched their personal choice, while for 24 others the choices differed. This shows that some players did indeed vote against their personal preferences; for the others, it is possible that their personal preferences happened to align with those of their assigned role.

The majority of participants personally preferred \textit{Wana’Like} (37 votes), ahead of \textit{Eye’Wana} (18) and \textit{Wana’Pass} (18). In contrast, at the conclusion of the game sessions, the city council's victory primarily went to \textit{Eye’Wana} (51 votes, three sessions won), followed by \textit{Wana’Like} (30 votes, two victories). This confirms that participants are capable of defending a solution on behalf of the role they play, even when it does not correspond to their personal opinion.

\subsection{Qualitative Analysis of Game Perception}
We also asked more open-ended questions about the players' feelings. The responses were provided as free-text, and we analysed the most frequent words within them.

Regarding the feelings at the conclusion of the game (57 responses), the most frequent word was "good" ("bien") (14 occurrences), followed by "interesting" (6) and "reflect" ("réfléchir") (4). Concerning what the players liked most about this activity (72 responses), the debate stood out (24 occurrences, including 6 times "debating", 4 "exchanges", and 4 "arguments"), as well as the act of playing a role within a group, with "group" (12) and "role" (9). High school students seem to greatly appreciate being able to discuss the health situation with their peers and expressed a desire to debate more often. The roles also allow everyone to gain perspective and express themselves freely without fear of judgment regarding their personal opinions.

There were 56 responses about negative points of the games. Various points emerged, although the word "nothing" ("rien") (7 occurrences) was cited most (indicating nothing negative to report). Several participants expressed a lack of time for the debates. Some regretted the limited amount of specific knowledge obtained about AI: it should be clarified in the introduction that this is not the goal of this game (its aim being to provide basic knowledge and a critical mindset regarding the adoption of AI-based technologies accessible to the general public, rather than teaching the technical details of how these technologies work, which would only be accessible with a certain level of computer science education). Others misunderstood the question and indicated which of the three solutions they liked the least. Some players felt frustrated that their preferred solution did not win. Finally, several expressed frustration because the solutions were judged all similar and all "bad", or because there was "no right answer."

This last point seems essential to us. Indeed, there is not always (and often not) a "right answer" to societal problems. What is important to teach students is how to look at these problems with a critical eye, to form their own opinions, and to defend their interests. The somewhat frustrating nature of the debate seems normal and even interesting to us; even if all participants have different opinions, the majority wins, and the others must abide by the collective choice. The debriefing is thus a vital part of such a serious game, as it allows for a discussion of the emotions felt during the game to gain perspective.

\subsection{Quantitative Analysis of the Game Subjective Impact}

We then sought to evaluate the impact of the game on the participants. Measuring this objectively is challenging, as the game does not provide traditional academic knowledge. The subject of AI is not part of the high school curriculum and therefore falls more under the category of general knowledge. The game objective is not so much to provide technical knowledge as to foster a certain critical mindset. Our questions therefore focused primarily on the players' subjective perceptions.
When asked if they thought they had learned anything during the activity, participants (54 responses) gave an average score of 6/10. This relatively low score clearly shows that the game does not really provide "hard skills" regarding AI, which may be disappointing, but rather "soft skills" or social competencies, such as the ability to debate, listen to others' arguments, and exercise critical thinking. The development of these skills is likely more difficult for high school students to assess than the acquisition of concrete knowledge.
We also asked them to judge how the game content connected with their studies. 29 of them considered the game to be complementary, 9 found it off-topic, 1 redundant, and 42 checked "other", among whom 7 specified that the subject was interesting but not part of their curriculum. This aligns with the game objective of providing a general AI culture to non-specialists.
Another question asked players if the game had made them think. The score here was better (average of 7/10, across 53 responses), confirming that the game's impact lies more in the reflections it triggered than in formal school learning. Encouraging students to reflect seems essential to us in the education of the citizens of tomorrow.
Finally, we asked participants what the game had made them want to do next (40 responses obtained). Among the proposed statements (multiple-choice questions), 18 checked "find out more about AI" and 20 others "continue the debate", which were our primary objectives. None checked "direct my studies toward AI", which could stem from the emphasis placed on the biases and dangers of AI rather than its technical functioning or advantages. Since the goal is not to turn students away from AI, care should be taken to better balance the discourse during the game.

Therefore, the debate game seems to have met our objectives, even if these are based on self-reported declarations. Since the questionnaires were anonymous and the sessions were organized in various institutions across the region, we are not in a position to ensure long-term follow-up to verify whether the students actually continued the debate or their search for information.

\subsection{Analysis of the Game Objective Impact}
To more objectively evaluate the game impact, we also asked participants what they had taken away from it. The responses were free-text, and 61 players responded. Several types of messages emerged: the necessary protection of privacy and personal data; the numerous biases, dangers, pitfalls, and disadvantages of AI, with words like "dangerous", "intrusive" or "surveillance"; a feeling of mistrust, or even fear ("creepy"); and the need to reflect, question oneself, and stay informed. These key messages correspond perfectly to the research interests of the two authors of this chapter who facilitated the sessions reported here. It is therefore crucial for the facilitators of this game to be aware of the messages they wish to convey to participants and of their own potential biases.

\subsection{Analysis of Game Biases}

Our objective through this activity is to raise citizen awareness and open a debate on the societal stakes of AI technologies used in our daily lives. The activity requires no prior knowledge of AI technologies, nor does it aim to provide any, even though technical and ethical aspects are intertwined. Certain biases may arise from our design of the activity. To clarify these, we collected feedback from researchers, teachers, and instructional designers during workshops and conferences.

The first point that emerges concerns the roles, which appear too stereotyped or unrepresentative of real population categories: political opinions, socio-professional categories, etc. Indeed, the roles are based more on age or geographical location. However, we aim to maintain non-politicized profiles in order to "cool down" the debate rather than heating it. Furthermore, the roles are solely intended to show a variety of viewpoints, without seeking to be exhaustive. Finally, our evaluation shows that the majority of players enjoyed playing the roles, succeeded in portraying them even if they did not identify with them, and benefited from the game regardless of the role played.

We have also been criticized for the competition between roles during the municipal debate to impose their preferred solution instead of fostering cooperation between roles to co-construct a common solution. Indeed, the activity is centered on this debate of ideas and forces everyone to argue their points in the face of differing perspectives. However, the adversarial debate could be extended by a complementary activity: the joint development of legislative specifications governing the use of the chosen solution to satisfy all citizens.

Some also regret the underlying technosolutionism of the activity, which forces a choice among the three proposed solutions. Unfortunately, this scenario is quite realistic: most students will have to live with numerous AI technologies and must therefore learn to weigh their benefits and risks. The game thus requires players to reflect on the positive and negative aspects of each solution before making a decision, never accepting them a priori, and to justify all their choices. These skills will be highly useful to them in their lives as citizens.

Finally, it should be noted that the proposed technologies only make sense within a certain legislative framework, as their consequences for individual liberties depend on how they are used. However, within the scope of the game, this legal context is intentionally kept somewhat vague to maintain a certain degree of generality. In our view, the same questions must be raised regardless of the context, because a technology implemented for a specific objective and within a given legal framework can then more easily be made permanent and extended, and the legislative framework itself can evolve. It is therefore important for citizens to ask questions, to be aware of the possible misuses of a technology, and to establish safeguards, such as a maximum duration of use. Thus, during the debriefing, the facilitators provide information on legal protections for users of AI technologies, in particular the right guaranteed by the General Data Protection Regulation (GDPR), but often ignored, to refuse to have an AI solution automatically make decisions concerning them.

\section{Practical Implementation Advice}

This debate game has been published on the \textit{L’Arbre des Connaissances} website since April 2022, as part of their series of debate games dedicated to AI. The various episodes are regularly downloaded and facilitated in classrooms by teachers. Implementation is straightforward thanks to the materials provided and the few use tips below.

\subsection{Target Audience}
\textit{L’Arbre des Connaissances} develops its activities specifically for high schools. However, we have also tested our debate-game with university students. It seemed poorly suited for Master level students who already possess specific knowledge about AI and expect more technical content. On the other hand, it works perfectly at the undergraduate level (Licence), where beyond the theme of AI, it primarily serves to provide skills in debating, argumentation, and critical thinking, all of which are highly useful for their studies regardless of their major. It can also be used to train non-computer science teachers. Finally, the theme can be adapted to foster reflection and debate among teachers regarding the growing role of AI in education.

\subsection{Setting Up the Activity}
To implement this game in the classroom, a time slot of approximately one and a half hours is required, with a class size of between 15 and 30 students. To facilitate discussions within the groups, the tables can be organized into five separate "islands" (group clusters). A large board will be needed to record the evaluations before the debate. Markers or chalk in several colours will be useful if multiple debate phases are planned, in order to visualize the evolution of the scores.

All necessary materials are freely available upon registration on the \textit{L’Arbre des Connaissances} website. They simply need to be downloaded and printed. The file contains cards describing the five roles, which are identical across all episodes in the series. One card should be provided per role, placed face down on the table of the relevant group. Assignment to groups is done randomly to prevent participants from choosing a group they identify with more closely.

The file also includes a sheet explaining the agenda for the city council, which is specific to the episode (in this case, dealing with an epidemic). This sheet highlights the important points to consider when evaluating the solutions. Only one copy is necessary, as the facilitator is responsible for reading and explaining it at the beginning of the session.

Finally, the file contains the solution sheets, which are also specific to each episode. There is one sheet per solution, providing Sowana marketing pitch, a few details on the AI techniques used in that solution, and a citizen's testimonial intended to raise some initial questions. These sheets inform players about each solution without being overly technical and suggest some limitations through a user testimonial to kickstart the internal group debate. Each solution sheet should be printed in five copies (one per group) and distributed one after the other, allowing time for internal debate on each solution before distributing the next one.

\subsection{Facilitation Guide}

The \textit{L’Arbre des Connaissances} website provides links to supplementary resources as well as a facilitator's guidebook intended for the teachers supervising the game. This guide contains useful definitions and more technical information to delve deeper into the concepts at hand; evaluation questions and scales to fuel the debate; information on similar real-world AI solutions to be presented during the debriefing; and links to additional documentation to explore the subject further.

The role of the facilitator is essential to the smooth running of the activity. They moderate the debate to ensure, as recommended by \cite{dieleman2006games}, a fair, calm, and balanced process. They may also be required to remind participants to speak strictly on behalf of their assigned role rather than expressing their personal opinions. The facilitator themself must also ensure they remain neutral and guide the debate without steering it; they must be aware of their own biases and the risk of communicating them to the players.

The objective of this game is to educate on the subject of AI from both a technical perspective (the general functioning of underlying algorithms) and an ethical one (the questions raised by its use). However, the mere act of playing and debating is not enough to achieve this goal. At the end of the activity, the facilitator is responsible for debriefing the game to allow for the assimilation of its messages. A proper debriefing is essential for learning through serious games (\cite{crookall2010serious}; \cite{dieleman2006games}; \cite{whalen2018all}) and serves several purposes: clarifying the concepts handled during the game session; linking the simplified concepts from the game to the complexity of the real world to ensure the transfer of learning to real-life situations; and sharing experiences, emotions, and reflections among participants. The debriefing must therefore be conducted by an informed moderator with a clearly defined pedagogical objective in mind.

In the context of this debate-game on the use of AI for health surveillance, the facilitator can show how the various solutions discussed during the game actually refer to real-world solutions already implemented in certain countries. These examples provide a very concrete illustration of the risks associated with shifting from the useful to the forced use of AI. They help participants grasp our key message: they are all affected by AI and its application.

\section{Conclusion}

In this chapter, we have presented a debate game designed to discuss with students the benefits and risks of AI technologies applied to a societal problem, illustrated here by the management of an epidemic. The specific theme is incidental, as the questions raised are far more general. Indeed, AI is becoming increasingly pervasive, and all citizens must be capable of making informed decisions about it. Furthermore, and even beyond the scope of AI, participants particularly appreciated being able to debate among their peers. The skills provided by this game involve argumentation, critical thinking, and listening to diverse viewpoints, essential "soft skills" for future citizens. In this regard, the principle of this debate-game could be extended to other themes and disciplines.

\small \enlargethispage{35pt}
\bibliographystyle{abbrv}

\normalsize


\end{document}